\documentclass[journal,draftcls,onecolumn,11pt,twoside]{IEEEtran}
\usepackage{cite}
\usepackage[latin1]{inputenc}
\usepackage[english]{babel}
\usepackage{amsmath,amsfonts,amssymb}
\usepackage{graphicx, epsfig}
\usepackage{subfigure}
\usepackage{array}
\usepackage{multirow}
\usepackage{mdwmath}
\usepackage{mdwtab}
\usepackage{algorithmic}
\usepackage{algorithm}
\usepackage{color}

\ifCLASSINFOpdf

\else

\fi

\begin{document}
%
\title{\huge Iterative Detection and Decoding Algorithms using LDPC Codes for MIMO Systems in Block-Fading Channels }


\author{Andre~G.~D.~Uchoa,~\IEEEmembership{Member,~IEEE,}
        Cornelius~Healy,~\IEEEmembership{Member,~IEEE,}
        and~Rodrigo~C.~de~Lamare,~\IEEEmembership{Senior~Member,~IEEE}
\thanks{A. G. D. Uchoa*, C. Healy and Rodrigo C. de Lamare* are with the Communications Research Group University of York, Heslington York, YO10 5DD, UK and CETUC/PUC-RIO*, Brazil e-mail: (andre.uchoa, cth503, delamare)@york.ac.uk.}
\thanks{This work was partially supported by CNPq (Brazil), under grant 237676/2012-5.}}

\maketitle

\begin{abstract}
We propose iterative detection and decoding (IDD) algorithms with
Low-Density Parity-Check (LDPC) codes for Multiple Input Multiple
Output (MIMO) systems operating in block-fading and fast Rayleigh
fading channels. Soft-input soft-output minimum mean-square error
receivers with successive interference cancellation are considered.
In particular, we devise a novel strategy to improve the bit error
rate (BER) performance of IDD schemes, which takes into account the
soft \textit{a posteriori} output of the decoder in a block-fading
channel when Root-Check LDPC codes are used. A MIMO IDD receiver
with soft information processing that exploits the code structure
and the behavior of the log likelihood ratios is also developed.
Moreover, we present a scheduling algorithm for decoding LDPC codes
in block-fading channels. Simulations show that the proposed
techniques result in significant gains in terms of BER for both
block-fading and fast-fading channels.
\end{abstract}


%
\IEEEpeerreviewmaketitle

\section{Introduction}

 {Modern wireless communication standards for cellular
and local area networks advocate the use of Low-Density Parity-Check (LDPC)
codes for high throughput applications \cite{wlan.80211ac}. Since
multiple-input multiple-output (MIMO) systems are often subject to multi-path
propagation and mobility, these systems are characterized by time-varying
channels with fluctuating signal strength. In applications subject to delay
constraints and slowly-varying channels, only limited independent fading
realizations are experienced \cite{rasmussen.10,mmhm,lcbd,gbd,mbthp,ccmmimo}. A
simple and useful model that captures the essential characteristics of such
scenarios is the block-fading channel \cite{rappaport,armo,sicdma}. A family of
LDPC codes called Root-Check codes were proposed in \cite{boutros.07} and can
achieve the maximum diversity of a block-fading channel when decoded with the
Belief Propagation (BP) algorithm. Recent LDPC techniques \cite{salehi.10,
iswcs.11, peg.comms.11, iswcs.12,dopeg,iswcs.13,vfap} that improve the coding
gain and have low-complexity encoding and reduced storage requirements have
been investigated.}

 { MIMO systems  can bring significant multiplexing
\cite{foschini.98, telatar.99} and diversity gains
\cite{alamouti.98, tarokh.99} in wireless communication systems. In
the block-fading channel the structure of the channel and the
degrees of freedom introduced by multiple antennas must be exploited
in order to appropriately design the receiver. Approaches to
receiver design include MIMO detectors \cite{did,vblast,
xia.08,df.99, df.05, poor.99, pingli.05, lamare.tcomms.08,
matsumoto.03, mimojio,peng.twc.11,tds,dfcc,mbdf}, decoding
strategies \cite{schedul.wesel} and iterative detection and decoding
(IDD) schemes \cite{poor.99,ldpc.cdma.mimo.2004}.
 {Among the most cost-effective detectors are the
successive interference cancellation (SIC) used in the Vertical Bell
Laboratories Layered Space-Time (VBLAST) systems \cite{vblast, xia.08} and
decision feedback (DF) \cite{df.99, df.05, poor.99, pingli.05,
lamare.tcomms.08, matsumoto.03, peng.twc.11,dfcc,jpais,tds,mfdf} techniques.
These suboptimal detectors can offer a good trade-off between performance and
complexity.} Prior contributions on IDD schemes include the seminal work of
Wang and Poor with turbo concepts \cite{poor.99} and the LDPC-based scheme
reported by Yue and Wang \cite{ldpc.cdma.mimo.2004}. In IDD schemes, the
decoder plays an important role in the overall performance and complexity. Vila
Casado and et. al. in \cite{schedul.wesel} have suggested that the use of
appropriate scheduling mechanisms for LDPC decoding can significantly reduce
the number of required iterations.
 {Prior work on MIMO detectors and IDD schemes have
dealt with quasi-static Rayleigh fading channels or fast Rayleigh fading
channels.} However, there are very few studies related to the case of
block-fading channels with MIMO systems.  {To the best of our knowledge, the
only study which discusses MIMO systems under block-fading channels is the work
by Capirone and Tarable \cite{mimobf.11}. They have shown that using Root-Check
LDPC codes with MIMO allows a system to achieve the desired channel diversity.}

{In contrast, in our work two key elements of an IDD system are
considered. First, by properly manipulating the log-likelihood
ratios (LLRs) at the output of the decoder and exploiting the code
structure we can obtain significant gains over standard LLR
processing for IDD schemes in block fading channels. Second, to
improve the overall performance we introduce a new scheduling
strategy for block-fading channels in IDD systems. The main
contributions of our work are the development of a novel IDD scheme
that exploits the code structure and a novel strategy for
manipulation of LLRs that improves the performance of MIMO IDD
systems in block-fading channels. In addition, we have also
developed a method of sequential scheduling to further improve the
performance of MIMO IDD systems in block-fading channels. The gains
provided by the proposed IDD scheme and algorithms do not require
significant extra computational effort or any extra memory storage.}

The rest of this paper is organized as follows. In Section II we
describe the system model. In Section III we discuss the proposed
log-likelihood ratio (LLR) compensation strategy. In Section IV we
introduce the proposed scheduling method. Section V analyzes some
aspects of the proposed techniques. Section VI depicts and discusses
the simulation results, while Section VII concludes the paper.

\section{System Model}

 {Consider a Root-Check LDPC-coded MIMO point-to-point
transmission system with $n_{tx}$ transmit antennas and $n_{rx}$ receive
antennas, where $n_{tx} \geq n_{rx}$.  {The system encodes a block of $L =
\frac{N}{m}$ symbols $\mathbf{s} = \left[s_{1}, s_{2}, \cdots, s_{L}
\right]^{T}$ from a constellation $\mathcal{A} = \left\lbrace a_{1}, a_{2},
\cdots, a_{C} \right\rbrace$, where $(\cdot)^{T}$ denotes the transpose,
$C=2^m$ denotes the number of constellation points and $m$ is the number of
bits per symbol, with a Root-Check LDPC encoder with rate $\frac{1}{F}$ for
each transmit antenna and obtains a block of $N$ encoded symbols $\mathbf{x} =
\left[x_{1}, x_{2}, \cdots, x_{N} \right]^{T}$.} At each time instant $t$, the
encoded symbols of the $n_{tx}$ antennas are organized into a $n_{tx} \times 1$
vector $\mathbf{x}[t] = \left[x_{1}[t], x_{2}[t], \cdots, x_{n_{tx}}[t]
\right]^{T}$ and transmitted over a block-fading channel with $F$ independent
fading blocks. The received signal is demodulated, matched filtered, sampled
and organized in an $n_{rx} \times 1$ vector $\mathbf{r}[t] = \left[ r_{1}[t],
r_{2}[t], \cdots, r_{n_{rx}}[t] \right]^{T}$ with sufficient statistics for
detection which is described by
\begin{equation} \label{eq:recsymbol}
 \mathbf{r}[t]= \sum_{k = 1}^{n_{rx}} \mathbf{h}_{k,f} \cdot x_{k}[t]+\mathbf{v}[t] =
 \mathbf{H}\mathbf{x}[t] + \mathbf{v}[t],
\end{equation}}
where the $n_{rx} \times 1$ vector $\mathbf{v}[t]$ is a zero mean
complex circular Gaussian noise with covariance matrix
$E\left[\mathbf{v}[t]\mathbf{v}^{H}[t]\right] =
\sigma_{\mathbf{v}}^{2}\mathbf{I}$, where $E[\cdot]$ stands for the
expected value, $(\cdot)^{H}$ denotes the Hermitian operator,
$\sigma_{\mathbf{v}}^{2}$ is the noise variance, $\mathbf{I}$ is the
identity matrix,  {$t=\{1,2,\cdots, \frac{L}{n_{tx}}\}$ is the time
index and $f=\{1,2,\cdots,F\}$ is the index corresponding to the
fading instants.} Moreover, $t$ and $f$ are related by $f=\lceil
F\cdot n_{rx} \cdot \frac{t}{L}\rceil$,  {where $\lceil \cdot
\rceil$ is a ceiling function}. In the case of fast fading we assume
that each received symbol will experience a distinct fading
coefficient, which means $F = L$. The uncoded symbol vector
$\mathbf{s}$ has a covariance matrix $E\left[ \mathbf{s}
\mathbf{s}^{H}\right] = \sigma_{\mathbf{s}}^{2} \mathbf{I}$, where
$\sigma_{\mathbf{s}}^{2}$ is the signal power.  {The model
(\ref{eq:recsymbol}) is used to represent the data transmission,
where each block of symbols is associated with a fading
coefficient.} For a given block, the encoded symbol vector
$\mathbf{x}$ is obtained by mapping $\mathbf{s}$ into coded bits and
forming the vector $\mathbf{x} = \left[x_{0}, \cdots, x_{j}, \cdots,
x_{n_{tx}\cdot m - 1} \right]^{T}$. The elements
$\mathbf{h}_{n_{rx}, n_{tx}}$ of the $n_{rx} \times n_{tx}$ channel
matrix $\mathbf{H}$ represent the complex channel gains from the
$n_{tx}$-th transmit antenna to the $n_{rx}$-th receive antenna.
 {In our paper, we define the signal-to-noise ratio
(SNR) as ${\rm SNR} = n_{tx}\cdot \frac{E_{s}}{R\cdot m \cdot
N_{0}}$.} An IDD scheme with a soft MIMO detector and LDPC decoding
is used to assess the performance of the system. The soft MIMO
detector incorporates extrinsic information provided by the LDPC
decoder, and the LDPC decoder incorporates soft information provided
by the MIMO detector. We call inner iterations the iterations done
by the LDPC decoder, and outer iterations those between the decoder
and the detector. In addition, in the decoder a novel scheduling
method is used for block-fading channels. The proposed scheduling
method combines the benefits of the Layered Belief Propagation (LBP)
and the Residual Belief Propagation (RBP) \cite{schedul.wesel}
algorithms as will be discussed in Section \ref{sec:idd_schedul}.
{In the IDD scheme, for the j-th code bit $x_{j}$ of the transmitted
vector $\mathbf{x}$ of each antenna, the extrinsic LLR of the
estimated bit of the soft MIMO detector is given by
\begin{equation} \label{eq:ext_bit}
l_{E}[x_{j}] = l_{C}[x_{j}] - l_{A}[x_{j}],
\end{equation}
where $l_{A}[x_{j}]$ is the \textit{a priori} LLR ($l_{A}[x_{j}] = 0$ at the
first iteration) of the bit $x_{j}$ computed by the LDPC decoder in the
previous iteration ($l_{C}[x_{j}] = 0$ at the first iteration) and
$l_{C}[x_{j}]$ is the \textit{a posteriori} LLR of the bit $x_{j}$ computed by
the soft MIMO detector. We have adopted in this work linear minimum mean square
error receive filters with SIC (MMSE-SIC) \cite{vblast} receivers. Other
detectors and receive filters can also be employed
\cite{wlmwf,wljio,mberjio,ccmjio,jidf,jio,stjio,mcg,ccmmwf,int,itic,stapjidf,vsscmv,jioccm}.}

\section{Proposed LLR Compensation Scheme}

We have investigated the performance of Root-Check LDPC codes in MIMO systems
with IDD schemes using MMSE-SIC \cite{vblast}. In particular, we have studied
numerous scenarios where Root-Check LDPC codes lose in terms of bit error rate
(BER) to the standard LDPC codes at high SNR.  {We have observed in simulations
that the parity-check nodes from Root-Check LDPC codes do not converge. In
particular, with Root-Check LDPC codes the LLRs exchanged between the decoder
and the detector degrade the overall performance.} To circumvent this, we have
adopted the use of controlled doping via high-order Root-Checks in graph codes
\cite{doping.boutros.11}. In our studies, the LLR magnitude of the parity check
nodes connected to the deepest fading always presented lower magnitude level
than the other parity check nodes. In contrast, for the case of standard LDPC
codes this magnitude difference has not been verified. For the case of
Root-Check LDPC codes, the difference in LLR magnitude (gaps) at the decoder
output for the parity check nodes has lead us to devise an LLR compensation
strategy to address these gaps. The gaps and the lower LLR magnitude for the
parity check nodes place the LLR values close to the region associated with the
non-reliable decision. {In addition, in an IDD process such values can cause
the detector to wrongly de-map the received symbols. Therefore, we have devised
an LLR processing strategy for IDD schemes in block-fading channels
(LLR-PS-BF).} {First, the \textit{a posteriori} LLRs generated by the soft MIMO
detector are organized in the N-dimensional vector $\mathbf{l}_{C} =
\left[l_{C}[x_1], l_{C}[x_2],\cdots, l_{C}[x_N]\right]$. Assuming that the
systematic symbols for a Root-Check LDPC code always converge to an LLR
magnitude greater than zero, we proceed to the following calculations:}
\begin{equation} \label{eq:alpha}
\alpha = \underset{1 \leq j \leq K} {\mathrm{max}}  (\vert l_C[x_j]
\vert) \text{ and } \beta = \underset{K+1 \leq j \leq N}
{\mathrm{max}} (\vert l_C[x_j] \vert),
\end{equation}
where $K$ is the length of the systematic bits.
 {We then compute $\gamma = \alpha - \beta$,
where $\gamma > 0$ due to the fact that the systematic nodes for a
Root-Check LDPC code always converge to an LLR magnitude greater
than zero. Once $\gamma$ is computed, we can generate a vector
$\mathbf{l}_{PA}$ described by
\begin{equation} \label{eq:lpa}
l_{PA}[j] = \vert l_{C}[x_j]\vert,~j = K+1, \cdots, N,
\end{equation}
which represents the positive magnitude of all parity-check  nodes.
We then calculate the vector $\mathbf{l}_{PS}$ as described by}
\begin{equation} \label{eq:lps}
l_{PS}[j] = \mathrm{sign} \left[ l_{C}[x_j] \right],~j = K+1, \cdots, N,
\end{equation} which corresponds to the signals of all parity-check nodes.
Furthermore, we obtain the vector $\mathbf{l}_{PT}$ as
\begin{equation} \label{eq:lpt}
\mathbf{l}_{PT} = \left( \mathbf{l}_{PA} + \gamma\right) \odot \mathbf{l}_{PS},
\end{equation}
where $\odot$ is the Hadamard product. The final step in the
proposed LLR-PS-BF algorithm is to generate the \textit{a
posteriori} LLRs to be used by the IDD scheme. Therefore, the
optimized vector of the \textit{a posteriori} LLRs is given by
\begin{equation} \label{eq:lcopt}
\mathbf{\tilde{l}}_{C} = \left[ l_{C}[x_{1}], \cdots, l_{C}[x_{K}], l_{PT}[x_{K+1}], \cdots, l_{PT}[x_{N}] \right].
\end{equation}
 {The aim of calculating $\mathbf{l}_{PT}$ is to
ensure that the LLRs of the parity-check nodes do not get close to
the region associated with non-reliable decisions. As a consequence,
the LLRs fed back to the detector will not deteriorate the
performance of the de-mapping operation. In the Appendix, we detail
how the proposed LLR-PS-BF compensation scheme works.}

We have carried out a preliminary study \cite{wcnc.14} where the LLR
compensation is a particular case of the one presented in this work.
In order to obtain the LLR-PS-BF scheme presented in \cite{wcnc.14}
we should set some different values. In particular, $\beta = 0$ and
$\mathbf{l}_{PA} = 0$ will lead to the same results presented in
\cite{wcnc.14}. It must be noted that every time the soft MIMO
detector generates an \textit{a posteriori} LLR $\mathbf{l}_{C}$ the
LLR-PS-BF compensation scheme must be applied when Root-Check LDPC
codes are used.  {The main purpose of applying the proposed
LLR-PS-BF compensation scheme is to enable convergence of the LLRs
to suitable values and preserve useful information in the
iterations.} Therefore, the LLRs exchanged between the decoder and
the detector will benefit from this operation. Consequently, a
better performance in terms of BER will result.

\section{Proposed IDD Scheme Based on Scheduling} \label{sec:idd_schedul}


The structure of the proposed LLR-PS-BF with the IDD scheme is
described in terms of iterations. In this work, we only consider the
use of SIC which outperforms the parallel interference cancellation
(PIC) detection scheme. When using SIC, the soft estimates of
${\mathbf r}[t]$ are used to calculate the LLRs of their constituent
bits. We assume that the $k$-th layer MMSE filter output $u_{k}[t]$
is Gaussian and the soft output of the SISO detector for the $k$-th
layer is written as \cite{lamare.tcomms.08}
\begin{equation} \label{eq:soft_out_siso_dect}
u_{k}[t] = \mathbf{V}_{k} x_{k}[t]+\epsilon_{k}[t],
\end{equation}
where $\mathbf{V}_{k}$ is a scalar variable which is equal to the
$k$-th layer's signal amplitude and $\epsilon_{k}[t]$ is a Gaussian
random variable with variance $\sigma_{\epsilon_{k}}^{2}$, since
\begin{equation} \label{eq:vk}
\mathbf{V}_{k}[t] = E \left[x_{k}^{*}[t] u_{k}[t]\right]
\end{equation} and
\begin{equation} \label{eq:sigma_ek}
\sigma_{\epsilon_{k}}^{2} = E \left[\vert
u_{k}[t]-\mathbf{V}_{k}[t]x_{k}[t] \vert^{2} \right].
\end{equation}
The estimates of $\mathbf{\hat{V}}_{k}[t]$ and
$\hat{\sigma}_{\epsilon_{k}}^{2}$ can be obtained by time averages
of the corresponding samples over the transmitted packet. After the
first iteration, the MMSE soft cancellation performs SIC by
subtracting the soft replica of Multiple Access Interference (MAI)
components from the received vector as
\begin{equation} \label{eq:soft_cancelation}
\mathbf{\hat{r}}_{k}[t] = \mathbf{r}[t] -
\sum_{j=1}^{k-1}\mathbf{h}_{j} \mathbf{\hat{x}}_{j}[t].
\end{equation}
The soft estimation of the $k$-th layer is obtained as $u_{k}[t] =
\boldsymbol\omega_{k}^{H} \hat{\mathbf{r}}_{k}[t]$, where the
$n_{rx} \times 1$ MMSE filter vector is given by
$\boldsymbol\omega_{k} = \left( \mathbf{H}_{k}\mathbf{H}_{k}^{H}
\boldsymbol\sigma_{v}^{2}\mathbf{I}\right)^{-1}\mathbf{h}_{k}$ and
$\mathbf{h}_{k}$ denotes the matrix obtained by taking the columns
$k, k+1, \cdots, n_{rx}$ of $\mathbf{H}$ and $\mathbf{\hat{r}}[t]$
is the received vector after the cancellation of previously detected
$k-1$ layers. where the soft output of the filter is also assumed
Gaussian. The first and the second-order statistics of the coded
symbols $\mathbf{\hat{x}}[t]$ are also estimated via time averages
of (\ref{eq:vk}) and (\ref{eq:sigma_ek}). We have developed our
proposed IDD scheme by applying scheduling methods for decoding LDPC
codes.  {Specifically, we have applied the Layered Belief
Propagation (LBP) scheduling method as described in
\cite{schedul.wesel} to evaluate the overall performance versus the
standard BP.} We have observed a performance loss for the scheduling
methods in the error floor region (high SNR region). To overcome
this problem we have applied our proposed LLR-PS-BF scheme. As a
result, the LBP has outperformed the standard BP as expected.

{Based on the result obtained by LBP we have applied the Residual Belief
Propagation (RBP) and the Node-Wise Belief Propagation (NWBP) to assess the
overall performance.}  {However, RBP and NWBP are outperformed by the standard
BP. The reason is that the block-fading channel imposes some constraints in
terms of LLRs received by the variable nodes. For practical purposes, let us
assume a block-fading channel with $F = 2$ fadings and that half of the
variable nodes have no channel information as the example given by Boutros
\cite[pp. 4, Fig. 10]{boutros.07}.} Furthermore, the idea of RBP and NWBP is to
prioritize the update of a specific message or check node with the largest
residual and then keep doing this in an iterative way. However, as soon as the
block fading channel affects the messages sent by $\frac{N}{2}$ variable nodes
to the check nodes, prioritizing such messages or nodes with no channel
information leads to a performance degradation. {Moreover, Gong and et.al. in
\cite{yigong.2011} have reported that all dynamic scheduling strategies only
concentrate on the largest residual when performing new residual computations.
Nonetheless, the existence of smaller residuals do not mean the algorithm in
the sub-graph of the Tanner graph has converged.}

The NWBP strategy has certain advantages over RBP because it
reinforces the root connections of a check node. It updates and
propagates simultaneously all the check-to-variable messages
$\mathbb{M}_{c_{i} \rightarrow v_{b}}$ that correspond to the same
check node $c_{i}$ as
\begin{equation} \label{eq:nwbp_mcivb}
\mathbb{M}_{c_{i} \rightarrow v_{b}}~:~\forall v_{b} \in \mathcal{N}(c_{i}),
\end{equation}
 {where $\forall v_{b} \in \mathcal{N}(c_{i})$ refers
to all variable nodes $v_{b}$ that belong to the set of check nodes
$\mathcal{N}(c_{i})$ that are connected to $v_{b}$.} Then, it
proceeds to calculate all the variable-to-check messages
$\mathbb{M}_{v_{b} \rightarrow c_{a}}$ that correspond to the same
variable node $v_{b}$ as
\begin{equation} \label{eq:nwbp_mvbci}
\mathbb{M}_{v_{b} \rightarrow c_{a}}~:~\forall c_{a} \in \mathcal{N}(v_{b}) \setminus c_{i},
\end{equation}
where $\mathcal{N}(v_{b}) \setminus c_{i}$ is the set of variable
nodes $v_{b}$ that are connected to $c_{a}$ except $c_{i}$. As a
result, NWBP will individually treat per iteration the check node
$c_{i}$ with the largest residual, which in the case of a
block-fading channel is not enough to gather all information
required by the root connections.  {However, we can address this if
at the beginning of each decoding iteration we calculate for each
check node the metric given by}
\begin{equation} \label{eq:checknode_metric}
\varphi_{c_{i}} = \max~r\left( \mathbb{M}_{c_{i} \rightarrow
v_{b}}\right)~:~\forall v_{b} \in \mathcal{N}(c_{i}),
\end{equation}
Following the example graph given in \cite[pp. 4, Fig.
10]{boutros.07}, we consider that the first half of the variable
nodes are under fading with $h_{1} = 1$ and the second half has no
channel information, i.e., $h_{2} = 0$, and $M_{CH} = \frac{N}{2}$
check nodes. Therefore, after $20$ inner iterations we can have the
following values:
\begin{eqnarray}  \label{eq:varphi_values}
\varphi_{c_{1, \cdots, \frac{M_{CH}}{2}}} & = & 0, \nonumber \\
\varphi_{c_{\frac{M_{CH}+1}{2}, \cdots, M_{CH}}} & \geq & 1.
\end{eqnarray}
Then, we can solve the block-fading problem by generating a queue
$Q$ of all $\varphi_{c_{i}}$ in a descending order from the largest
to the smallest to obtain the corresponding indexes of the check
nodes as
\begin{equation} \label{eq:queue_ordering}
Q = \left[i_{1}, i_{M_{CH}} \right] \therefore \lbrace \varphi_{c_{a}} \in \mathbb{N} : \varphi_{c_{i_{1}}} > \varphi_{c_{a}} > \varphi_{c_{i_{M_{CH}}}} \rbrace.
\end{equation}
Therefore, in a pre-defined order based on the queue $Q$, an
iteration consists of the sequential update of all variable to check
messages $\mathbb{M}_{v \rightarrow c}$ as well as all the check to
variable messages $\mathbb{M}_{c \rightarrow v}$. This approach is
called Residual LBP (RLBP).



Therefore, if we adopt a strategy like RLBP it will lead to a
prioritization, at each iteration, of the largest to the smallest
check-to-variable residual being updated and propagated. As a
result, we still have a performance degradation compared to the
standard LBP. It turns out that, as discussed in \cite{yigong.2011},
the smaller residuals of the sub-graph on the Tanner graph do not
necessarily indicate convergence. We have then devised a dynamic
scheduling strategy which overcomes the problems caused by a
block-fading channel. The proposed scheduling strategy, called
Residual Ordered LBP (ROLBP), alternates at each decoding iteration
between two different strategies.  {For every other iteration the
ROLBP strategy requires the computation of the check nodes metric
(\ref{eq:checknode_metric}) and ordering (\ref{eq:queue_ordering})
while RLBP requires this for every iteration. The ROLBP technique
can be described by the following calculations:}

First, initialize all $\mathbb{M}_{c\rightarrow v} = 0$ and all
$\mathbb{M}_{v_{j}\rightarrow c_{i}} = C_{v_{j}}$, where $C_{v_{j}}$
is the channel information LLR of the variable node $v_{j}$. Then,
compute all the residuals of the messages as
\begin{equation} \label{eq:residuals}
r(\mathbb{M}_{c\rightarrow v}), ~\textrm{ generate} ~Q,
\end{equation}
where $Q$ is the list of residuals in descending order. We then
proceed to the calculation of $\Xi$ as
\begin{equation} \label{eq:Xi}
\Xi = \left\{
  \begin{array}{lr}
   Q(1), \cdots, Q(M_{CH}), & \textrm{if the iteration is odd}\\
   1, \cdots, M_{CH}, & \textrm{if the iteration is even}
  \end{array}
\right..
\end{equation}
For each $i \in \Xi(1),\cdots, \Xi(M_{CH})$ calculate:
\begin{equation} \label{eq:mvkci}
\forall c_{i} \in {\cal N} (v_{j}) \to \textrm{generate and
propagate } \mathbb{M}_{v_{k} \rightarrow c_{i}}
\end{equation}
\begin{equation} \label{eq:mcivk}
\forall v_{k} \in {\cal N} (c_{i}) \to \textrm{generate and
propagate } \mathbb{M}_{c_{i} \rightarrow v_{k}}
\end{equation}
\begin{equation} \label{eq:rmcvq}
\textrm{Update and compute} \to \textrm{ All }
r(\mathbb{M}_{c\rightarrow v}) \textrm{ regenerate } Q
\end{equation}
Finally, if the decoding stopping rule is not satisfied then
recalculate all the equations from (\ref{eq:residuals}) up to
(\ref{eq:rmcvq}). Again returning to the example given in \cite[pp.
4, Fig. 10]{boutros.07}, the values of $\varphi_{c_{i}}$ for ROLBP
throughout the iterations are:
\begin{equation}  \label{eq:varphi_rolbp}
\varphi_{c_{1, \cdots, M_{CH}}} \geq 0,
\end{equation}
which results in a scheduling method that decreases the
prioritization as seen in (\ref{eq:varphi_values}).
 {By adopting this strategy we ensure that ROLBP
outperforms both the standard BP and RLBP algorithms. The reason is
that we give enough information to the root connections and avoid
the values for $\varphi_{c_{i}}$ as in (\ref{eq:varphi_values})
which cause a degradation in performance of Root-Check based LDPC
codes.} The pseudo-code is described in Algorithm
\ref{alg:schedul_idd_alg1}.
 {The computational complexity of the decoding
algorithms depends on the variable node degree $d_v$ and the check
node degree $d_c$ . The number of edges in the Tanner graph is
$\epsilon = d_v N_{VN} = d_c N_{CN}$, where $N_{VN}$ is the number
of variable nodes and $N_{CN}$ is the number of check nodes.
 {In terms of complex multiplications, one $\epsilon$
update of BP corresponds to $d_c N_{CN}/4$ operations, $d_c
N_{CN}(1+(d_v-1)(d_c-1))/4$ operations for NWBP, $d_c N_{CN}/4$
operations for LBP, $d_c N_{CN}/2$ operations for RLBP, and $1.5 d_c
N_{CN}/2$ operations for ROLBP. The most complex decoding algorithm
is NWBP, which is followed by RLBP, the proposed ROLBP algorithm, BP
and LBP.}


\section{Simulations} \label{sec:simul}

The bit error rate (BER) performance of the proposed LLR-PS-BF with
a SIC IDD scheme is compared with Root-Check LDPC codes and LDPC
codes using a different number of antennas. It must be remarked that
our proposed LLR-PS-BF scheme can be applied to other types of IDD
schemes \cite{peng.twc.11}. Both LDPC codes used in the simulations
have block length $N = 1024$ for all rates. The maximum number of
inner iterations was set to $20$ and a maximum of $5$ outer
iterations were used.  { The Root-Check LDPC codes require less
iterations than standard LDPC codes for convergence of the decoder
(inner iterations) \cite{salehi.10,peg.comms.11}. Using Root-Check
LDPC codes in IDD schemes reduces the need for inner iterations,
whereas the number of outer iterations remains at five. We have used
codes with rates $1/2$ and $1/4$ for the sake of transmission
efficiency and because they can be of practical relevance. Rates
lower than $1/4$ are not attractive in terms of efficiency.}  We
considered the proposed algorithms and all their counterparts in the
independent and identically-distributed (i.i.d) block fading channel
model. The coefficients are taken from complex circular Gaussian
random variables with zero mean and unit variance. The modulation
used is QPSK. The SNR at the receiver is calculated as $SNR_{RCV} =
\frac{1}{2\cdot \sigma_{\epsilon_{k}}^{2}}$ which is based on
equation (\ref{eq:sigma_ek}).

 {In Fig. \ref{fig:2x2subf2} the results for a
point-to-point $2 \times 2$ MIMO system, block-fading channel with
$F = 2$ fadings and code rate $R = \frac{1}{2}$ are presented along
with an illustration of the computational complexity of the decoding
algorithms in complex multiplications. The proposed LLR-PS-BF scheme
with Root-Check LDPC codes using the ROLBP strategy outperforms BP
by about $1$ dB in terms of SNR for the same BER performance. When
we compared the LLR-PS-BF with a Root-Check LDPC scheme with both
using ROLBP, LLR-PS-BF has a gain of up to $2$ dB in terms of SNR
for the same BER performance. The gain of the ROLBP algorithm alone
is also up to $2$ dB in SNR for the same BER performance. The
complexity of the ROLBP algorithm is higher than that of the
standard BP and the LBP algorithms but lower than the RLPB and NWBP
algorithms.}

Fig. \ref{fig:4x4subf2} presents the results for a point-to-point $4
\times 4$ MIMO system, block-fading channel with $F = 2$ fadings and
code rate $R = \frac{1}{4}$. On average, all Root-Check based codes
using LLR-PS-BF compensation outperform the standard LDPC codes for
all decoding strategies. In addition, ROLBP outperforms BP by about
$1.25$ dB. ROLBP with LLR-PS-BF outperforms standard LDPC codes with
BP by up to $1.5$ dB in terms of SNR for the same BER performance.

Fig. \ref{fig:2x2sufast} shows the outcomes for a point-to-point $2
\times 2$ MIMO system, fast-fading channel and code rate $R =
\frac{1}{2}$. As the BER performance for standard LDPC codes using
different decoding strategies has lead to the same performance, we
have plotted only one curve to represent BP, LBP and ROLBP. The
gains of the proposed LLR-PS-BF IDD scheme using ROLBP are about $1$
dB with respect to standard LDPC codes. Furthermore, at low SNR the
LLR-PS-BF scheme with ROLBP has outperformed LBP by about $1.5$ dB
in terms of SNR.  {The scenarios with $F=L/2$ or $F = L/4$ cases can
be addressed by using Root-Check LDPC codes with $F = 2$ and the
proposed LLR compensation scheme. In particular, the design of
Root-Check LDPC codes for $F = L/2$, $F = L/4$ or other $F$ is
unnecessary as the Root-Check LDPC code with $F = 2$ is able to
capture the advantages for a wide range of $F$.}

\section{Conclusion}

In this paper, we have presented an IDD scheme for MIMO systems in block-fading
channels.  {Furthermore, we have proposed the ROLBP scheduling algorithm  for
the proposed IDD scheme and studied different scheduling strategies. The
proposed algorithms have resulted in a gain of up to $2$ dB for a
point-to-point $2 \times 2$ MIMO system and up to $1.5$ dB for a $4\times 4$
MIMO system in a block-fading channel with $F = 2$. For the case of a $2\times
2$ MIMO system over fast-fading the proposed LLR-SP-BF IDD scheme has obtained
a gain of up to $1.5$ dB. The proposed algorithms are suitable for MIMO systems
with users that experience high throughput rate and slow changes in the
propagation channel.} In such scenarios, the symbol period is much
smaller than the coherence time. 


\appendix[LLR-PS-BF Mathematical Analysis] \label{ap:a}

Mathematically speaking, we can interpret the LLR-PS-BF compensation
scheme as a modification made by two functions $f[\mathbf{l}_{C}]$
and $g[\mathbf{l}_{C}]$. Given $\mathbf{l}_{C}$, an input vector of
length $N$, we consider $K = \frac{N}{2}$ which is true for code
rate $R = \frac{1}{2}$. First, the aim of $f[\mathbf{l}_{C}]$ is to
obtain a real value $\Delta \in \Re^{+}$. Therefore, we have
\begin{equation*} \label{eq:fx}
\Delta = f[\mathbf{l}_{C}] =
  \begin{array}{lr}
    \max(l_{C}) & , l_{C}[1], \cdots, l_{C}[K]\\
  \end{array}.
\end{equation*}
Finally, the discrete signal $\mathbf{l}_{C}$ is processed by
$g[\mathbf{l}_{C}]$ to generate the compensated version of
$\mathbf{l}_{C}$ called $\tilde{\mathbf{l}}_{C}$. Therefore,
$g[\mathbf{l}_{C}]$ is defined as
\begin{equation*} \label{eq:gx}
g[\mathbf{l}_{C}] = \left\{
  \begin{array}{lr}
    l_{C} & , l_C[1], \cdots, l_{C}[K]\\
    l_{C}+\frac{l_{C}}{\vert l_{C}\vert}\cdot \Delta & , l_{C}[K+1], \cdots, l_{C}[N]
  \end{array}
\right.,
\end{equation*}
where $\frac{l_{C}}{\vert l_{C}\vert}$ is the sign of $l_{C}$ and
$\tilde{\mathbf{l}}_{C} \Leftarrow g[l_{C}]$. To further understand
how the functions $f[\mathbf{l}_{C}]$ and $g[\mathbf{l}_{C}]$ act in
the input vector $\mathbf{l}_{C}$ we provide an example in Fig.
\ref{fig:appendix} for a vector $\mathbf{l}_{C}$ with $N = 1024$ and
$K = 512$. We only show the parity-check LLRs ($K
> 512$). On the left had side of Fig. \ref{fig:appendix} we have the
non-optimized version of $\mathbf{l}_{C}$ while on the right hand
side we depict the compensated $\tilde{\mathbf{l}}_{C}$. As we can
see from Fig. \ref{fig:appendix}, for the non-optimized vector
$\mathbf{l}_{C}$ some of the parity-check LLRs tend to the region
associated with non-reliable decisions while the compensated version
$\tilde{\mathbf{l}}_{C}$ places the parity-check LLRs farther from
such region.

\bibliographystyle{IEEEtran}
%

%
\bibliography{IEEEabrv,long_paper_sc}


\begin{figure}[!htb]
 \centering
\resizebox{120mm}{!}{
\includegraphics{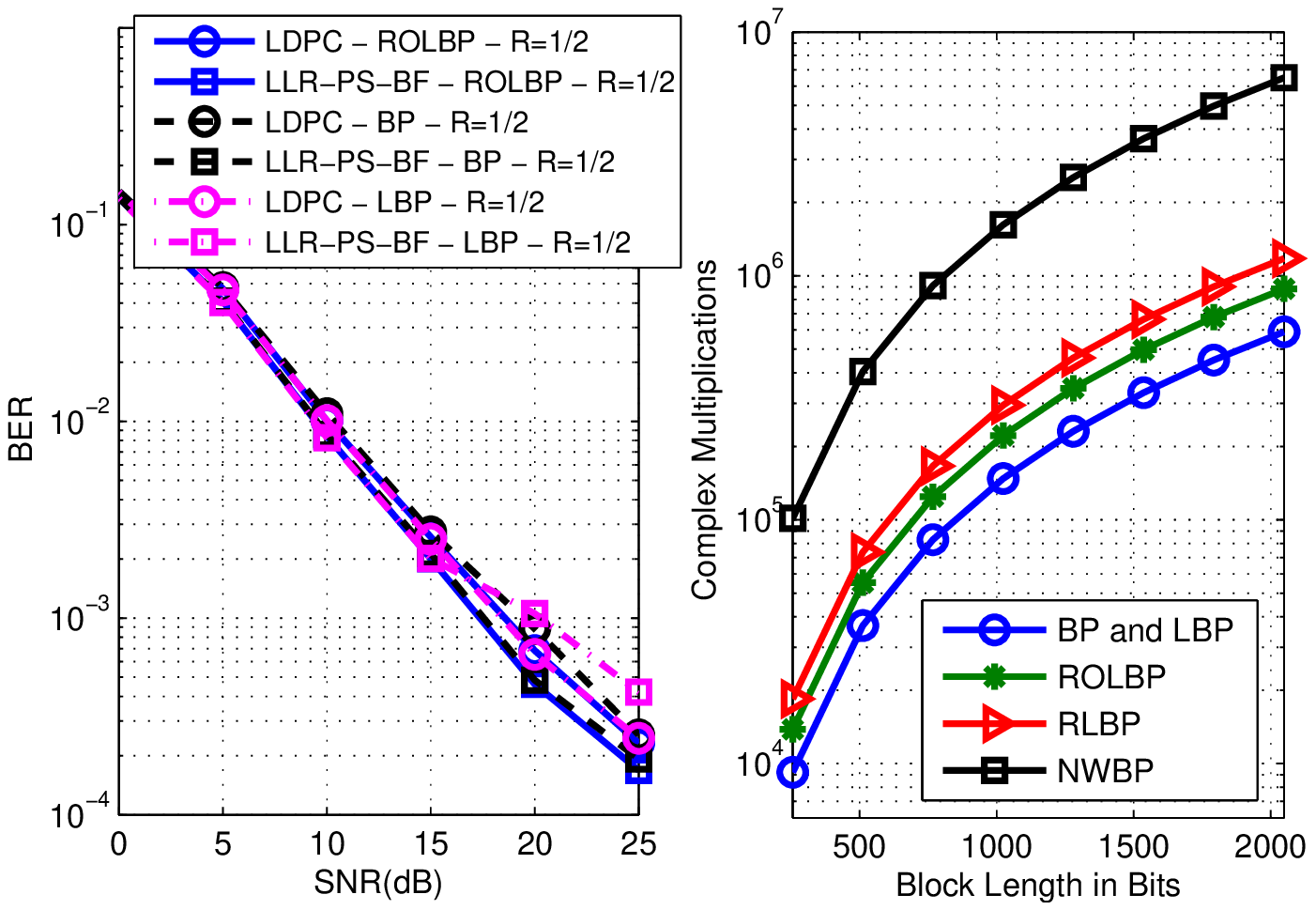}
} \vspace{-1em}\caption{BER performance of LLR-PS-BF with Root-Check
LDPC versus LDPC code both codes are rate $R = \frac{1}{2}$ and
block length $N = 1024$. The decoding strategies considered are BP,
LBP and ROLBP and the computational complexity is expressed in
complex multiplications. A point-to-point MIMO system with $2 \times
2$ configuration in a block-fading channel with $F = 2$, QPSK
modulation, $5$ outer iterations and $20$ inner iterations is used.}
\label{fig:2x2subf2}
\end{figure}

\begin{figure}[!htb]
 \centering
\resizebox{120mm}{!}{
\includegraphics{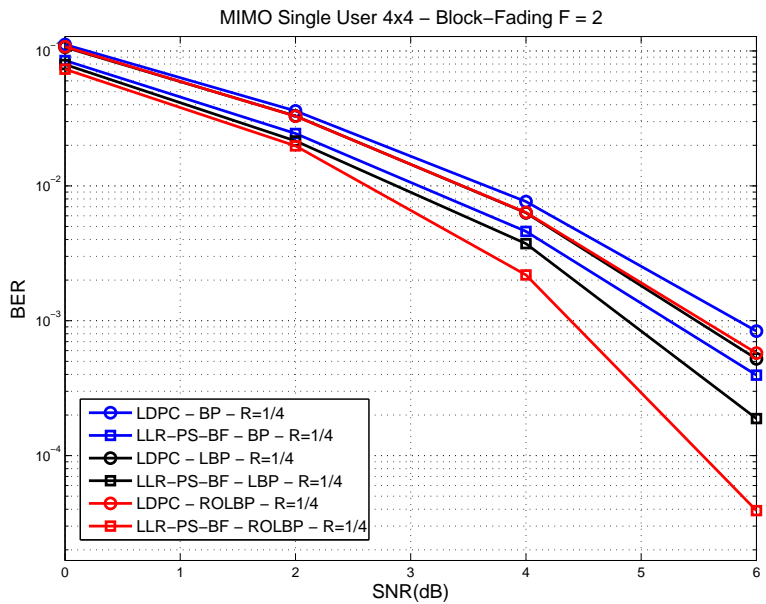}
}\caption{BER performance of LLR-PS-BF with Root-Check LDPC versus
LDPC code. The codes have rate $R = \frac{1}{4}$ and block length $N
= 1024$. The decoding strategies considered are BP, LBP and ROLBP. A
point-to-point MIMO system in a $4 \times 4$ configuration in a
block-fading channel with $F = 2$, QPSK modulation, $5$ outer
iterations and $20$ inner iterations is employed.}
\label{fig:4x4subf2}
\end{figure}

\begin{figure}[!htb]
 \centering
\resizebox{120mm}{!}{
\includegraphics{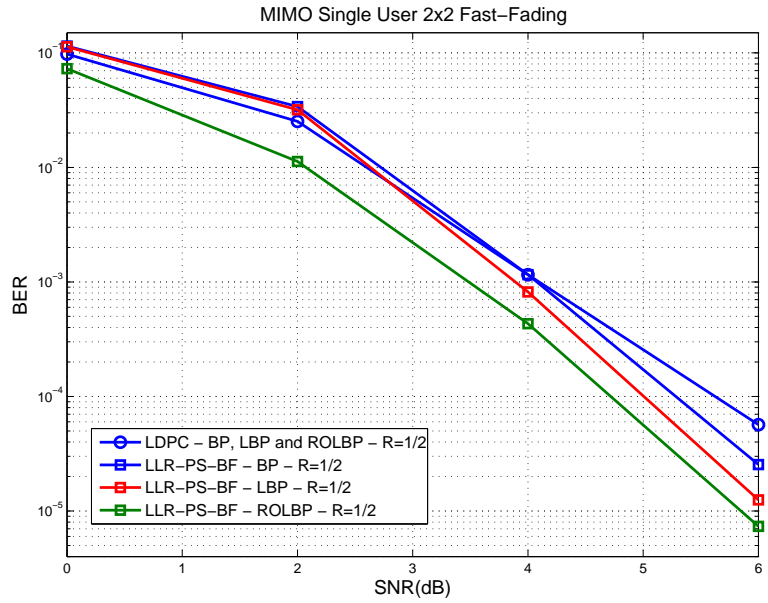}
}\caption{BER performance of LLR-PS-BF with Root-Check LDPC versus
LDPC code. The codes have rate $R = \frac{1}{2}$ and block length $N
= 1024$. The decoding strategies considered are BP, LBP and ROLBP. A
point-to-point MIMO system with a $2 \times 2$ configuration in a
fast-fading channel is considered, QPSK modulation, $5$ outer
iterations and $20$ inner iterations is used.} \label{fig:2x2sufast}
\end{figure}

\begin{figure}[!htb]
 \centering
\resizebox{120mm}{!}{
\includegraphics{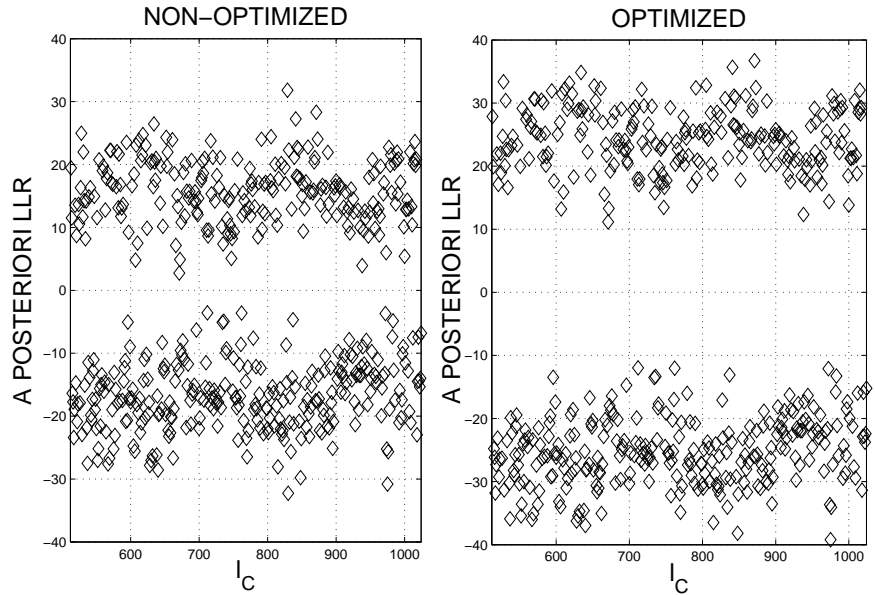}
}\caption{An example of the optimization of $\mathbf{l}_{C}$ made by
the proposed LLR-PS-BF compensation scheme. For the case of length
$N = 1024$, $K = 512$ and code rate $R = \frac{1}{2}$.}
\label{fig:appendix}
\end{figure}

\begin{algorithm}[!htb]
{\small
 \caption{Proposed LLR-SP-BF Scheduling IDD Scheme}
 \label{alg:schedul_idd_alg1}
\algsetup{
linenosize=\small,
linenodelimiter=.
}
\begin{algorithmic}[1]
\STATE \textbf{Require}: $\mathbf{r}[{t}]$, $\mathbf{H}$,
$\sigma_{v}^{2}$, $\mathbf{l}_{A}$ \textit{a priori} information,
$\mathit{TI}$.
\medskip
\FOR{$\mathrm{l_{0}} = 1 \to TI~\lbrace \mathrm{Turbo~Iteration}\rbrace$}
\medskip
    \STATE Calculate MMSE filter
    $\boldsymbol{w}_{k} = \left(\mathbf{H}_{k,f}\mathbf{H}_{k,f}^{H} + \frac{\sigma_{v}^{2}}{\sigma_{s}^{2}} \mathbf{I} \right)^{-1} \mathbf{h}_{k,f}$
\medskip
    \STATE \textbf{Detection Scheme - SIC} \\

    $\mathbf{\hat{r}}_{k}[t] = \mathrm{Perform-SIC}( \mathbf{{r}}[t], \mathbf{H}, \sigma_{v}^{2}, \mathbf{w}_{k})$, perform the MMSE SIC detection scheme for each $k$-th layer.\\
\medskip
    \STATE \textbf{Obtain The Extrinsic Bit LLR}
\medskip
    \STATE \textbf{First}: Determine $\sigma_{\epsilon_{k}}^{2}$ based on the best channel realization by means of calculating:
    $\delta_{f} = \underset{1\leq f \leq F}{\operatorname{arg~max}}\vert \det(\mathbf{h}_{k,f})\vert$, where $\delta_{f}$ is the index of $f$ which $\vert \det(\mathbf{h}_{k,f})\vert$ has the maximum value.
\medskip
    \STATE Therefore, $\mathbf{V}_{k}[t]$ and $\sigma_{\epsilon_{k}}^{2}$ must be calculated
    where the fading happens at index $\delta_{f}$. This is unique for block-fading channels,
    other types of channels do not require these additional steps. Then, the extrinsic LLR
    is obtained as: \\
    $l_{E}[x_{j}] = l_{C}[x_{j}] - l_{A}[x_{j}]$
\medskip
    \STATE \textbf{LDPC Decoding}
\medskip
    \IF{Using Scheduling}
\medskip
        \STATE Do the decoding with equations from (\ref{eq:residuals}) up to (\ref{eq:rmcvq});
\medskip
    \ELSE
\medskip
        \STATE Decode using standard belief propagation;
\medskip
    \ENDIF
    \STATE Obtain the a posteriori LLR  $\mathbf{l}_{C}$ of the soft MIMO detector.
\medskip
    \IF{$\mathrm{LDPC} = \mathrm{RootCheck}$}
\medskip
        \STATE \textbf{Apply the proposed LLR-PS-BF scheme equations (\ref{eq:alpha}) up to (\ref{eq:lcopt})}
\medskip
        \STATE Calculate the extrinsic information $l_{E}[x_{j}]$ based on $l_{C}[x_{j}]$ to be sent to the decoder.
\medskip
    \ELSE
\medskip
        \STATE Calculate the extrinsic information $l_{E}[x_{j}]$ based on $l_{C}[x_{j}]$ to be sent to the detector.
\medskip
    \ENDIF
\medskip
\ENDFOR
\end{algorithmic}
}
\end{algorithm}

\end{document}